\documentclass[fleqn,usenatbib,useAMS]{mnras}

\usepackage{ae,aecompl}
\usepackage{graphicx}
\usepackage{amsmath,amsfonts,amssymb}
\usepackage[T1]{fontenc}

\newcommand{\nuc}[2]{${}^{#2} \rm #1$}
\newcommand{\km}{{\, \rm km}}

\newcommand{\K}{ {\,\rm K} }
\newcommand{\Ms}{M_{\odot}}
\newcommand{\nue}{\nu_{\rm e}} 
\newcommand{\nueb}{{\bar \nu}_{\rm e}} 

\title[Impact of asymmetric neutrinos on nucleosynthesis]
{The impact of asymmetric neutrino emissions on nucleosynthesis in core-collapse supernovae}

\author[S. Fujimoto and H. Nagakura]{Shin-ichiro Fujimoto$^{1}$
\thanks{E-mail: fuji@kumamoto-nct.ac.jp},
Hiroki Nagakura$^{2}$
\\
$^{1}$National Institute of Technology, Kumamoto College, Kumamoto 861-1102, Japan
\\
$^{2}$Department of Astrophysical Sciences, Princeton University, Princeton, NJ 08544, USA\\
}

\date{Accepted XXX. Received YYY; in original form ZZZ}

\pubyear{2019}

\begin{document}
\label{firstpage}
\pagerange{\pageref{firstpage}--\pageref{lastpage}}
\maketitle

\begin{abstract}
We investigate the impact of asymmetric neutrino emissions on the explosive nucleosynthesis in neutrino-driven core-collapse supernovae (CCSNe).
We find that the asymmetric emissions tend to yield larger amounts of proton-rich ejecta (electron fraction, $Y_e > 0.51$) in the hemisphere of the higher $\nue$ emissions,
meanwhile neutron-rich matter ($Y_e < 0.49$) are ejected in the opposite hemisphere of the higher $\nueb$ emissions.
For larger asymmetric cases with $\ge 30\%$, 
the neutron-rich ejecta is abundantly produced, in which there are too much elements heavier than Zn compared to the solar abundances.
This may place an upper limit of the asymmetric neutrino emissions in CCSNe. The characteristic features are also observed in elemental distribution;
(1) abundances lighter than Ca are insensitive to the asymmetric neutrino emissions:
(2) the production of Zn and Ge is larger in the neutron-rich ejecta even for smaller asymmetric cases with $\le 10\%$.
We discuss these observational consequences, which may account for the (anti-)correlations among asymmetries of heavy elements and neutron star kicks in supernova remnants (SNRs).
Future SNR observations of the direct measurement for the mass and spatial distributions of $\alpha$ elements, Fe, Zn and Ge
will provide us the information on the asymmetric degree of neutrino emissions.
\end{abstract}




\section{Introduction}\label{sec:intro} 

Core-collapse supernova explosion (CCSN) marks the death of massive stars, in which the heavy elements in the order of solar mass are ejected into the interstellar medium.
The explosive nucleosynthesis during the development of explosion and the subsequent neutron star cooling phases are particularly important for the production of heavier elements.
Importantly the direct measurements of some of the heavy elements in young supernova remnants (SNRs) are now possible
and offer valuable insights to understand the explosion mechanism and the possible link with the neutron star (NS) kick~\citep{2006A&A...457..963S, 2006A&A...453..661K, 2013A&A...552A.126W, 2017ApJ...842...13W}.

It is now established that multi-dimensional(D) hydrodynamic instabilities are one of the key ingredients in the explosion mechanism.
They disorganize the post-shock accretion flows including the envelope of the NS, which also produce temporal variations and global asymmetries in the neutrino emissions.
This may also impact on the explosive nucleosynthesis, which has been investigated based on results of 2D simulations of CCSNe~\citep{2005ApJ...623..325P, 2006ApJ...644.1028P, 2011ApJ...738...61F, 2011ApJ...726L..15W, 2013ApJ...767L..26W, 2013ApJ...774L...6W, 2017ApJ...843....2H, 2018JPhG...45a4001E, 2018ApJ...852...40W}.

More recently, the coherent asymmetric neutrino emissions have been witnessed in some CCSN simulations with detailed neutrino transport; for instances, lepton-emission self-sustained asymmetry appears in some 3D CCSN simulations among different groups~\citep{2014ApJ...792...96T, 2018ApJ...865...81O, 2019MNRAS.482..351V}
and large asymmetric neutrino emissions associated with the NS kick are observed in a latest axisymmetric CCSN simulation with full Boltzmann neutrino transport~\citep{2019arXiv190704863N}.
In this {\it Letter}, 
we examine the impact of the asymmetric neutrino emissions on the explosive nucleosynthesis by axisymmetric hydrodynamic simulations with approximate neutrino transport and by post-processing calculation of the abundance evolution of SN ejecta with a nuclear reaction network combined with the tracer particle method.
In this study, we focus on a representative SN progenitor; 19.4$\Ms$ with the solar metallicity~\citep{2002RvMP...74.1015W}.
Based on our results, we discuss the observational consequences of asymmetric neutrino emissions such as correlations of asymmetric distributions of heavy elements, ejecta morphology and NS kick in SNRs.

This paper is organized as follows. In section 2, we will presents results of 2D axisymmetric simulations of CCSN explosions.
In section 3, we present our results of nucleosynthesis and discuss observational implications in SNRs of our results.
Finally, we will summarize our conclusion in section 4.

\section{Aspherical CCSN explosion}\label{sec:SN explosion} 

We employ two-independent numerical codes in our CCSN simulations; one of which is an open-source spherically symmetric CCSN scheme, namely "GR1D"~\citep{2015ApJS..219...24O}, 
and the other is based on another open-source hydrodynamic scheme, namely "ZEUS-2D"~\citep{1992ApJS...80..791S, 2006ApJ...641.1018O, 2007ApJ...667..375O}, with incorporating required CCSN physics such as neutrino-matter interactions and neutrino transport.
We use the former for the simulation up to $\sim 100$ms after the bounce including the collapsing phase. We then switch to axisymmetric simulations with dynamically-unimportant velocity perturbations.
Most parts of the numerical method are the same as those used in~\citet{2011ApJ...738...61F}, in which we solve axisymmetric hydrodynamic equations with taking into account the feedback from neutrino-matter interactions.
The neutrino distribution function, which is required in the feedback from neutrinos to matter, is computed by a simplified $\nu$ transport model, namely "light-bulb $\nu$ transport"~\citep{2006ApJ...641.1018O}, 
in which neutrinos are emitted from a sphere with thermal distributions.

In this study, we assume that the neutrino temperatures are spherical symmetric
\footnote{We checked the average energy of neutrinos in some CCSN simulations with full Boltzmann neutrino transport, 
and confirmed that its asphericity is much smaller than that in the energy-flux.}
but the luminosities are aspherical;
\begin{eqnarray}
L_{\nue} &=& L_{\nue, \rm ave} (1 +m_{\rm asy} \cos \theta), \\
L_{\nueb} &=& L_{\nueb, \rm ave} (1 -m_{\rm asy} \cos \theta),
\end{eqnarray}
where $L_{\nue}$ and $L_{\nueb}$ are luminosity of $\nue$ and $\nueb$ and $L_{\nue, \rm ave}$ and $L_{\nueb, \rm ave}$ are their angle averages, respectively.
Here the evolution of $L_{\nue, \rm ave}$, $L_{\nueb, \rm ave}$ and neutrino temperatures
are evaluated with a $\nu$-core model from the mass accretion rate at an inner edge of the computational domain of our 2D simulations,
as in systematic studies on CCSN nucleosynthesis in spherical symmetry~\citep{2012ApJ...757...69U, 2016ApJ...821...38S} but with some modifications 
(see details in Fujimoto et al. (2019) in preparation).
For instances, $L_{\nue, \rm ave}$ and $L_{\nueb, \rm ave}$ were set to be constant in time and $m_{\rm asy} = 0$ in the previous work~\citep{2011ApJ...738...61F}.
We have tuned two parameters of the $\nu$-core model so that the explosion of the $19.4\Ms$ progenitor for $m_{\rm asy} = 0\%$ (spherical $\nu$ emission) reproduces 
observations of SN1987A, or the explosion energy $E_{\rm exp} \sim 10^{51} \rm erg$ and the ejected mass of \nuc{Ni}{56}, $M$(\nuc{Ni}{56}), of $\sim (0.07-0.08) \Ms$.
We set $m_{\rm asy}$ to be 0\%, 10/3\%, 10\%, 30\%, and 50\%\footnote{$m_{\rm asy} \sim 10 \%$ has been observed in the recent multi-D CCSN simulations (see, e.g., \citet{2014ApJ...792...96T}).}.

\begin{figure}
  \begin{center}
   \includegraphics[scale=1.0]{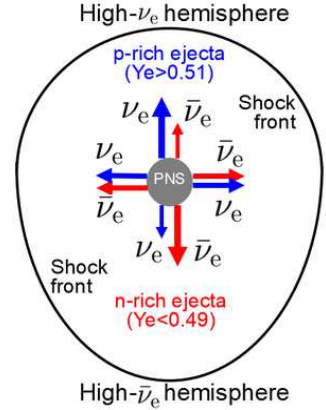}
  \end{center}
  \caption{
  Schematic picture of CCSN explosion with asymmetric $\nue$ and $\nueb$ emission. 
  The asymmetry leads to the compositional differences in SN ejecta, which become $n$($p$)-rich in the high-$\nue$ (high-$\nueb$) hemisphere 
  due to the enhancement of $\nue$ ($\nueb$) and the decline of $\nueb$ ($\nue$).}

  \label{fig:schematic}
 \end{figure}

We have performed the simulations for the progenitor of 19.4$\Ms$ up to $\sim$ 1.2s after the core bounce, 
in which shock fronts for all models have reached to a layer with $r = 10,000\km$ in almost all directions by the end of the simulations.
We confirm that the explosion is highly aspherical and $l=1$ and $2$ modes are dominant
as shown in \citet{2006A&A...453..661K, 2006A&A...457..963S, 2007ApJ...667..375O, 2011ApJ...738...61F}.
The aspherical neutrino emission makes ejecta in the high-$\nue$ hemisphere proton($p$)-rich due to enhanced $\nue$-absorptions on neutron($n$)
and declined $\nueb$-absorptions on $p$,
while the enhanced $\nueb$ and the declined $\nue$ emission results in $n$-rich ejecta in the high-$\nueb$ hemisphere (Fig. \ref{fig:schematic}).

\begin{figure}
  \begin{center}
   \includegraphics[scale=0.65]{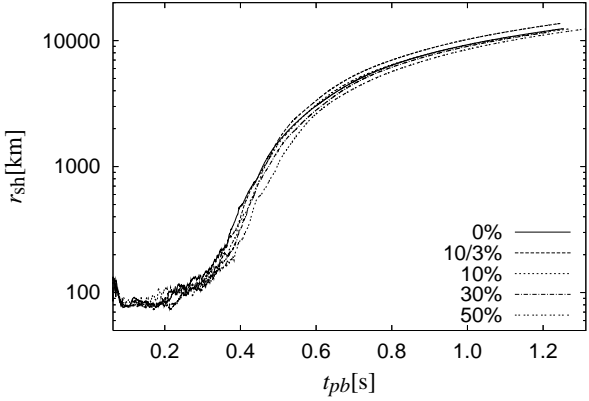}
  \end{center}
  \caption{Evolution of average shock radius, $r_{\rm sh}$, for cases with $m_{\rm asy}=$ 0\%(solid line), 10/3\%(dashed line), 10\%(dotted line), 30\%(dash-dotted line), and 50\%(double dotted line) as a function of the time from the core bounce $t_{\rm pb}$.
  }
  \label{fig:shock radii}
 \end{figure}

In our models, properties of the CCSN explosion weakly depend on $m_{\rm asy}$; 
the evolution of shock radii are similar (Fig. \ref{fig:shock radii}) and
$E_{\rm exp}$ and $M$(\nuc{Ni}{56}) are (0.89-1.17)$\times 10^{51} \rm erg$ and (5.2-8.1)$\times 0.01 \Ms$, respectively.
It should be noted, however, the asymmetric neutrino emissions may be associated with the shock morphology in reality (see e.g., \citet{2014ApJ...792...96T}).
This issue can be only addressed by CCSN simulations with more consistent treatments of neutrino-radiation hydrodynamics
such as ~\citet{2018ApJ...854..136N} and \citet{2019arXiv190704863N}.
We postpone these studies to a future paper.

\section{Abundances of SN ejecta}\label{sec:abundances}

For the nucleosynthesis computations, Lagrangian thermodynamic histories are computed by a tracer particle method~\citep{1997ApJ...486.1026N, 2010MNRAS.407.2297S},
in which the particles evolve in accordance with the fluid-velocity while storing their physical quantities such as density, temperature, and electron fraction.
At the beginning of the simulations, 6,000 tracer particles are distributed in the regions from $1,000$km to $10,000$km or an O-rich layer with being weighted to the mass in the shell.
By virtue of the adaptive particle mass distributions, the highest resolution in the particle mass is $\sim 10^{-4} \Ms$ in this study.
Note that 6,000 particles are sufficient to investigate the impact of asymmetric neutrino emissions in CCSN.
Indeed, in our previous study~\citep{2011ApJ...738...61F}, we checked the sensitivity of nuclear abundances to the particle numbers, 
and found that the difference of the ejected mass of heavy nuclei between 3,000 and 6,000 particles cases is less than $\sim$ 1\%.
For the initial condition, we take the abundance data of 20 nuclei from the result of stellar evolution in~\citet{2002RvMP...74.1015W}.


In our models, more than 3,500 particles are ejected by explosions.
The nuclear abundances produced by the explosive nucleosynthesis and their masses in the ejecta are estimated in the region $r_{\rm cc} \le 10,000\km$,
where the peak temperature is higher than the required temperature to produce elements heavier than C.
For this reason our nucleosynthesis computations are conducted only for the tracer particles which are located $r \le r_{\rm cc}$ at the initial condition
\footnote{The mass in the region $r > r_{\rm cc}$ for the 19.4 $\Ms$ progenitor is $13.0\Ms$.}.
We compute nucleosynthesis for 2,488 nuclides from $n$, $p$ to Nd by employing the same network as that in \citet{2007ApJ...656..382F}.
Note that, for the ejecta with higher temperature than $9 \times 10^9 \K$, 
the chemical composition of the ejecta is set to be that in nuclear statistical equilibrium.
Note also that the number of nuclides in this study is less than that in the previous study ($\sim 4,000$ nuclides),
but we improved some neutrino interactions for He and nuclei from C to Kr as in ~\citet{2011ApJ...738...61F}.
Albeit possessing $Y_e$ data from the hydrodynamic simulations,
we recompute $Y_e$ following the weak interactions employed in the nuclear reaction network, and we adopt them in our network computations.

\begin{figure}
  \includegraphics[scale=1.1]{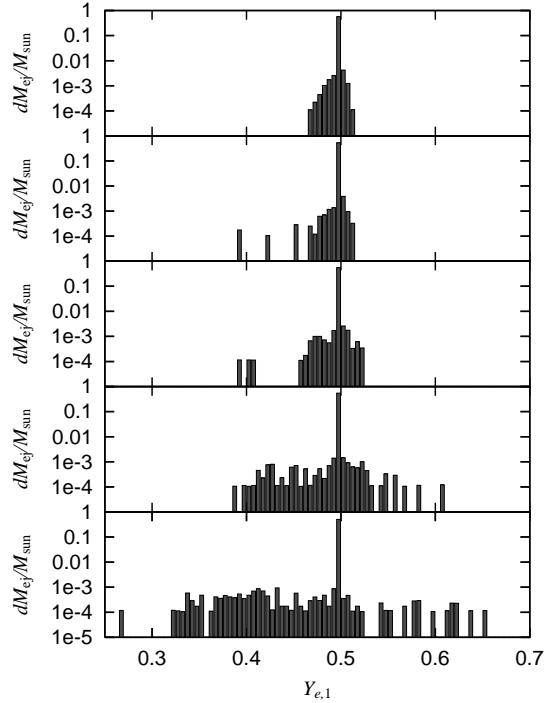}
  \vspace*{-30pt}
  \caption{Mass profiles of $dM_{\rm ej}$ in $Y_{e,1}$ of the inner ejecta ($r_{\rm cc} \le 10,000 \km$)
 for cases with $m_{\rm asy}=$ 0\%, 10/3\%, 10\%, 30\%, and 50\% in panels from top to bottom, respectively.}
  \label{fig:dist-ye-ww19.4}
\end{figure}

Below, we present the main results of our nucleosynthetic computation.
Figure \ref{fig:dist-ye-ww19.4} shows mass profiles of $dM_{\rm ej}$ in $Y_{e,1}$ of the ejecta from the inner region ($r_{\rm cc} \le 10,000\km$)
for cases with $m_{\rm asy}=$ 0\%, 10/3\%, 10\%, 30\%, and 50\%. 
Here $Y_{e,1}$ is the electron fraction evaluated when the temperature is equal to $10^9 \rm K$ during the ejection
and $dM_{\rm ej}$ is a mass of ejecta integrated with a bin of $dY_{e,1} = 0.005$.
We find that for larger $m_{\rm asy}$, amounts of both $p$- and $n$-rich ejecta become larger with wider range of $Y_{e,1}$.
This is mainly due to the excess of $\nue$ or $\nueb$ absorptions by asymmetric neutrino emissions.

\begin{figure}
 \includegraphics[scale=1.1]{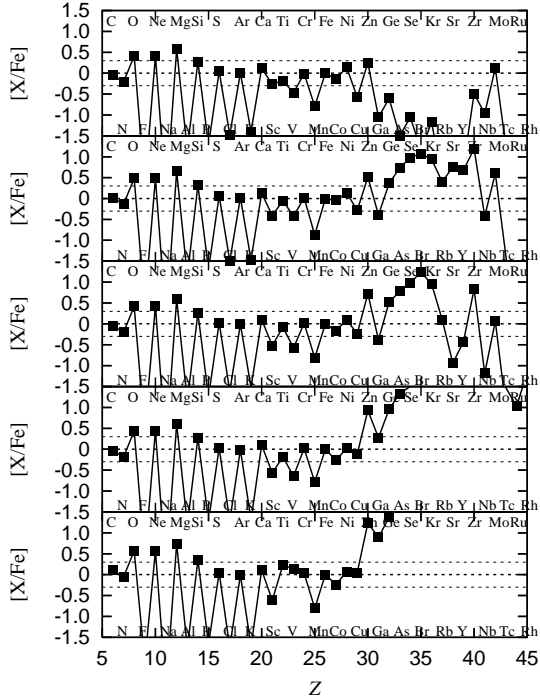}
 \vspace*{-30pt}
 \caption{
 [X/Fe] of all ejecta for cases with $m_{\rm asy}=$ 0\%, 10/3\%, 10\%, 30\%, and 50\% in panels from top to bottom.
 Cases with large asymmetry ($m_{\rm asy} =$ 30\% (fourth panel) and 50\% (bottom panel)), produce too much elements with $Z \ge 30$.}
 \label{fig:xfe-ww19.4}
\end{figure}

Figure \ref{fig:xfe-ww19.4} shows the composition of the ejecta for our models
in terms of [X/Fe]~\footnote{[A/B] = $\log \left[ (X_{\rm A}/X_{\rm A, \odot})/(X_{\rm B}/X_{\rm B, \odot})\right]$,
where $X_{\rm i}$ and $X_{\rm i,\odot}$ are a mass fraction of element $\rm i$ and its solar value~\citep{1989GeCoA..53..197A}, respectively.} 
of the ejecta as a function of an atomic number $Z$ for a model with $m_{\rm asy}=$ 0\%, 10/3\%, 10\%, 30\%, and 50\%.
Here masses of an element of the ejecta are evaluated with the sum of those in the inner ejecta ($r_{\rm cc} < 10,000\km$)
and those of ejecta from the outer layers ($\ge 10,000 \rm km$) with the progenitor abundances.
We find that abundances of the $p$-rich ejecta are similar to those in ejecta with $Y_e \sim 0.49-0.51$.
This is attributed to the fact that the $\nu p$ process is weak and $rp$ process hardly happen 
due to the low entropies and long expansion timescales in the ejecta, which results in small proton to seed ratios, as in \citet{2018ApJ...852...40W}. On the other hand, 
we find that the $n$-rich ejecta have a largely different composition from the ejecta with $Y_e \sim 0.49-0.51$ and have abundant elements with $Z \ge 29$,
which are abundantly produced through quasi and nuclear statistical equilibrium;
the composition is not largely changed through reactions other than $\beta$ decays even after the break of the equilibrium~\citep{2018ApJ...852...40W}.
We note that dependences of the ejecta composition on $Y_{e,1}$ are similar to those in \citet{2018ApJ...852...40W}.
Elements lighter than Al are chiefly synthesized in the progenitor and abundances of elements lighter than Ca are independent to $m_{\rm asy}$,
because of similar $E_{\rm exp}$ and amounts of ejecta with $Y_{e,1} \sim 0.5$ (Fig. \ref{fig:dist-ye-ww19.4}),
even for cases with large asymmetric $\nu$ emission ($m_{\rm asy} \ge 30\%$).
For cases with small $\nu$ asymmetry ($m_{\rm asy} =$ 10/3\% and 10\%), elements with $Z \ge 29$ 
can be abundantly produced in the ejecta with some overproduction of elements ($Z=$ 34-37 and $40$),
while models with large asymmetry ($m_{\rm asy} =$ 30\% and 50\%) produce too large elements with $Z \ge 30$ compared with those in the solar system,
due to too large $n$-rich ejecta (Fig. \ref{fig:dist-ye-ww19.4}).
Although the detailed comparison to the solar abundance requires considering the progenitor dependence and taking the IMF-average, 
such an extreme overproduction can not be easily diminished.
We thus conclude that $m_{\rm asy} \ge $ 30\% in $\nu$ luminosity is not favorable to yield the solar abundances from CCSNe.

\begin{figure}
 \includegraphics[scale=0.65]{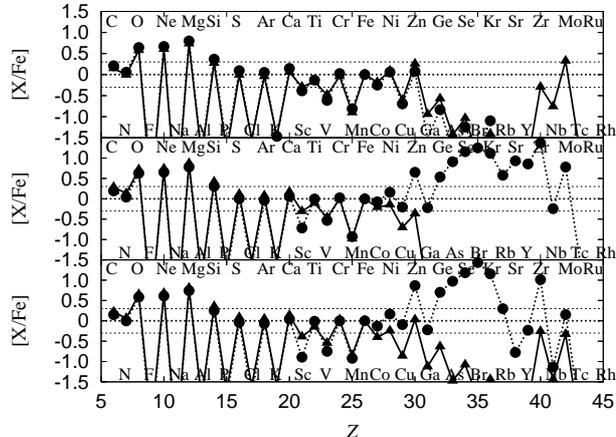}
 \vspace*{-10pt}
 \caption{
 [X/Fe] of ejecta in the high-$\nue$ (solid lines) and high-$\nueb$ (dotted lines) hemispheres 
 for $m_{\rm asy}=$ 0\% (top), 10/3\% (middle), and 10\% (bottom).
 Compositional differences between the high-$\nue$ and high-$\nueb$ hemispheres are prominent for Sc and elements with $Z \ge 29$
 for cases of aspherical $\nu$ emission.
 }
 \label{fig:xfe-ww19.4-ns}
\end{figure}
Figure \ref{fig:xfe-ww19.4-ns} shows [X/Fe] of ejecta in the high-$\nue$ (solid lines) hemisphere and high-$\nueb$ (dotted lines) one,
where $n$-rich ejecta ($Y_{\rm e,1} < 0.49$) are abundantly located, for cases with $m_{\rm asy}=$ 0\% (top panel), 10/3\% (middle panel), and 10\% (bottom panel).
We note that abundances in the high-$\nueb$ hemisphere are very similar to those of all the ejecta (Fig. \ref{fig:xfe-ww19.4}).
For the case with spherical $\nu$ emission (top panel), compositional differences between the northern- and southern- hemisphere is small.
Note that in the spherical case we make a geometrical distinction of northern and southern hemispheres instead of high-$\nue$ and high-$\nueb$ hemispheres.
On the other hand, the differences in Sc and elements with $Z \ge 29$ are prominent for cases with aspherical $\nu$ emission (middle and bottom panels).

In this study, we have adopted the limited number of nuclei as the progenitor composition.
To clarify effects of odd-$Z$ and $s$-process elements synthesized in the outer layers of the 19.4$\Ms$ progenitor, which are not included in our mass estimate,
we evaluate masses of Sc, Cu, Zn, Ga, and Ge in the outer layers ($r_{\rm cc} > 10,000\km$) of a SN progenitor of 19.0$\Ms$, 
whose detailed composition are evaluated for nuclei more than 1,200 just before the core collapse~\citep{2002ApJ...576..323R} 
and are likely to be comparable to those for the 19.4$\Ms$ progenitor.
When $\nue$ and $\nueb$ are spherically emitted, 
the differences of the above elements between northern- and southern- hemispheres possibly disappear,
since the elements are more massive in the outer layers of the progenitor.
For $m_{\rm asy} = $ 10/3\% and 10\% cases, 
the differences of Sc, Cu, and Ga are likely to be covered with larger masses of these elements in the outer layers.
It should be noted, however, that the differences are still prominent in masses of Zn and Ge, 
even if the elements are taken into account in the outer layers of the progenitor.

Finally we apply our results to the observational consequences in SNRs.
As indicated by some theoretical studies, the asymmetric neutrino emissions may be correlated with the ejecta morphology and NS kick, 
in which $\nueb$ is higher in the hemisphere of stronger shock expansion (opposite direction to the NS kick if the mechanism is the hydrodynamic origin) 
and $\nue$ is stronger in the opposite hemisphere (see, e.g., \citet{2014ApJ...792...96T} and \citet{2019arXiv190704863N}). 
If this is true, the asymmetric distributions of heavy elements found in the present study can be observed with the correlation with the NS kick.
We speculate that the distribution of elements would have the following characteristics;
abundances of elements lighter than Ca are insensitive to the ejecta morphology and NS kick direction, but Zn and Ge would be abundant in the opposite side to the NS kick.
Indeed, a recent observation of asymmetric motion of center of mass of $\alpha$ elements~\citep{2018ApJ...856...18K} may be relevant to our findings.
Note also that observations of abundances of $\alpha$ elements (O, Ne, Mg, Si, and S), and Fe in some SNRs have been proposed to estimate the mass of a SN progenitor~(e.g., \citet{2007ApJ...671.1717T, 2018ApJ...863..127K}),
and we find that their total abundances are insensitive to the asymmetric neutrino emissions.
Thus, the estimate could not be modified even for asymmetric neutrino emissions,
although the systematic study is indispensable in order to assess the impact more quantitatively.

We also note that the ejected mass of Zn found in the present study is not subtle, indeed, it is comparable to that of Cr and even higher than that of Mn, 
while the line emissions of the latter two elements from SNRs were observed by Suzaku X-ray satellite~(e.g., \citet{2013ApJ...766...44Y}).
X-Ray Imaging and Spectroscopy Mission (XRISM),
which will be launched at 2021, may detect them with distinguishing Zn-lines from others in young SNRs due to the high energy resolution in a X-ray calorimeter,
which is very similar to that in the Hitomi X-ray observatory.
It should be noted that the last Hitomi X-ray mission showed the ability of the high sensitivity of detecting faint line emissions of Zn from Perseus cluster~\citep{2019PASJ...71...50T}.
More quantitative assessment of the detectability requires detailed computations of the line emissions with multiple progenitors, which will be done in our forthcoming paper.

\section{Conclusion}\label{sec:conclusion}

We have examined explosive nucleosynthesis in CCSNe of the 19.4$\Ms$ progenitor under asymmetric and anti-correlated $\nue$ and $\nueb$ emission, 
assuming a dipolar, angular dependence (Eqs. 1 and 2) and varying magnitudes of the asymmetry, $m_{\rm asy}$, from 0\% (spherical emission) to 50\%.
Due to the dipolar and anti-correlated angular dependence of the $\nu$ luminosities,
SN ejecta become $p$-rich and $n$-rich in the high-$\nue$ and high-$\nueb$ hemispheres, respectively.
We also found that elements heavier than Cu ($Z \ge 29$) are abundantly produced in the $n$-rich ejecta (Fig. \ref{fig:xfe-ww19.4}).
The asymmetry ($m_{\rm asy} \ge 30\%$) leads to too much elements heavier than Zn ($Z \ge 30$) (e.g., bottom panel in Fig. \ref{fig:xfe-ww19.4}), 
which may imply that such a large asymmetry does not appear in reality. 
On the other hand, small asymmetry ($m_{\rm asy} \le 10\%$)
leads to produce elements heavier than Cu in the n-rich ejecta, which are deficit in the case with spherical $\nu$ emissions.

We also apply our results to the observational consequences of SNRs, 
in which Zn and Ge would be abundant in the opposite side of NS kick if the kick correlates with asymmetric $\nu$ emissions, 
which are seen in some recent multi-D simulations with detailed $\nu$ transport.
Our finding indicates that the correlation between the asymmetric distributions of heavy elements and NS kick may be 
not only due to the aspherical shock expansion~\citep{2017ApJ...837...84J} but also asymmetric $\nu$ emissions.
We also speculate that the observation of abundances of $\alpha$ elements, Fe, Zn, and Ge may give us information of asymmetric degree of $\nu$ emissions.

In the present study, we focus on a single representative SN progenitor. Although the qualitative trend may be the same in different progenitors, 
amounts of $p$-rich and $n$-rich ejecta depend not only on $m_{\rm asy}$ but also on the neutrino luminosities and the velocities of the ejecta.
Thus, the systematic study of the progenitor dependence with more detailed neutrino transport should be investigated for more quantitative arguments.
We also speculate that the asymmetric neutrino emissions from the PNS core potentially impact on $r$-process nucleosynthesis 
in ejecta from the core ($\le 50\,\km$), which were not included in the present computation.
This is also on the to-do-list of future investigations and the result will be published elsewhere.

\vspace*{-20pt}

\section{Acknowledgements}
We thank S. Katsuda for a helpful discussion about observational aspects of Zn in a SNR.
We also thank the anonymous referee for detailed comments that helped us to improve our manuscript.
This work is partly supported by JSPS KAKENHI Grant Number 25400281.



\vspace*{-10pt}


\input{ms.bbl}







\bsp	
\label{lastpage}
\end{document}